\documentstyle[12pt,cite]{article}

\def\Journal#1#2#3#4{{#1} {\bf #2}, #3 (#4)}

\def\NPB{{\em Nucl.Phys.}B}
\def\PLB{{\em Phys.Lett.}B}
\def\PRL{\em Phys.Rev.Lett.}
\def\PRD{{\em Phys.Rev.}D}

\def\JETP{\em JETP}

\def\PAN{\em Phys.At.Nucl.}

\newcommand{\eq}[1]{\begin{equation}#1\end{equation}}

\begin{document}






\topskip 2cm
\begin{titlepage}
\begin{center}

{\large\bf Parametric Resonance Amplification 
of Neutrino Oscillations in 
Electromagnetic Wave with Varying Amplitude \ \ \ \ \ \ \ \ \ \ \ \ \ \ \ \ \ \ 
and "Castle Wall" Magnetic Field}\\
\vspace{2.5truecm}
{\large M.S. Dvornikov,
A.I. Studenikin\footnote{\normalsize E-mail: studenik@srdlan.npi.msu.su}}
\vspace{.5cm}

{\sl Department of Theoretical Physics, Moscow State University,

119899 Moscow,Russia}

\vspace{2.5cm}
\vfil
\begin{abstract}
{Within the Lorentz invariant formalizm for description of neutrino 
evolution in electromagnetic fields and matter we consider neutrino 
spin oscillations in the electromagnetic wave with
varying amplitude and in "castle wall" magnetic field.  
It is shown for the first time that the parametric resonances of neutrino 
oscillations in such systems can occur.}
\end{abstract}
\end{center}
\end{titlepage}

\eject




Experimental studies of
solar, atmospheric and reactor neutrinos over the past several years
provide almost certain indications that neutrinos oscillate, have masses
and mix. These new properties of neutrinos, if confirmed by better
statistics in the proceeding and forthcomming
experiments, will require a significant departure from
the Standard Model. It is well known [1] that massive neutrinos can have
nonvanishing magnetic moment. For example, in the Standard Model supplied with
$SU(2)$-singlet right-handed neutrino the one-loop radiative correction
generates neutrino magnetic moment which is proportional to neutrino mass.
There are plenty of models [2] which predict much large magnetic moment for
neutrinos.

In the light of new developments in neutrino physics, to get better
understanding of the electromagnetic properties of neutrinos
is an important task. In general, there are two aspects of this problem.
The first one is connected with values of neutrino electromagnetic form - 
factors.
The second aspect of the problem implies consideration of influence of
external electromagnetic fields, which could be presented in various
environments, on neutrino possessing non - vanishing electromagnetic moments.
The most important of the latter are the magnetic and electric dipole moments.

The majority of the previously performed studies of neutrino conversions
and oscillations in electromagnetic fields deal with the case of
constant transversal magnetic field or constant transversal twisting
magnetic fields (see references [1-18] of paper [3]). Recently we have
developed [3-5] the Loretz invariant formalism for description of
neutrino spin evolution that enables one to consider neutrino oscillations
in the presence of an arbitrary electromagnetic fields. Within the
proposed approach it becomes possible to study neutrino spin evolution in
an electromagnetic wave and the new types of resonance in the neutrino
oscillations in the wave field and in some other combinations of fields have
been predicted. The latest application [6] 
of this formalism enables us to
consider neutrino oscillations in electromagnetic fields 
in moving and polarised matter and show that matter effects on
neutrino oscillations in the case of relativistic motion of matter 
could sufficiently depend on matter motion.

The aim of this paper is to continue the study of the neutrino spin
oscillations in different electromagnetic fields and consider the cases 
of electromagnetic wave with periodically varying amplitude (see also [7])
and 
periodic "castle - wall" magnetic field. It will be shown that under 
certain conditions the enhancement of neutrino oscillations could be 
achieved due to the parametric resonance effect.
 
The possibility of the parametric resonance of neutrino flavour oscillations
in matter with periodic variation of density was considered previously in [8].
It should be also pointed here that the conditions for a total neutrino
flavour conversion in a medium consisting of two or three constant density
layers were derived in [9]. For the recent discussion on the physical
interpretations of these two mechanisms of increasing of neutrino conversion
see
refs.[9-11]. 

First we consider
the case when the case of electromagnetic 
wave,  
with the amplitude of which being not constant but periodically varying
in time, and show (see also ref.[7])
that under certain conditions
the parametric resonance of neutrino oscillations can occur in such a system.
Note that in order to investigate this phenomenon we have to use the Lorentz
invariant formalisms for neutrino evolution that have been developed in [3-5].

Let us consider evolution of a system $\nu =(\nu_{+},\nu_{-})$ composed of
two neutrinos of different helicities in presence of a field of circular
polarised electromagnetic wave with periodically varying amplitude. Here
neutrinos $\nu_{+}$ and $\nu_{-}$ correspond to positive and negative helicity
states that are determined by projection operators
\eq{P_{\pm}={{1}\over{2}}
\left(1\mp {{({\vec \sigma}{\vec p})}\over{|\vec p|}}
\right).}
In general case it is important to distinguish helicity states
$\nu=(\nu_{+},\nu_{-})$ and chirality states $\nu=(\nu_{L},\nu_{R})$.
The latter are determined by the projection operators $P_{L,R}=(1\mp
\gamma _{5})/2$.
The evolution of such a system is given [3-5] by the following
Schr$\ddot o$dinger type equation

\eq{i{{\partial \nu}\over{\partial t}}=H\nu, \ \ \
H={\tilde \rho}\sigma_{3}+E(t)(\sigma_1 \cos \psi -
\sigma_2 \sin \psi ),}
$${\tilde \rho}=-{{V_{eff}}\over{2}}+{\Delta{m}^{2}A \over 4E}, \ \ \
E(t)=\mu B(t)(1-\beta\cos \phi).$$
Here the three parameters, $A=A(\theta)$ being a function of vacuum mixing
angle, $V=V(n_{eff})$ being the difference of neutrino effective
potentials in matter, and $\Delta m^{2}$ being the neutrino masses
squired difference depend on the considered type of neutrino conversion
process. The electromagnetic field is determined in the laboratory frame of
reference by its frequency, $\omega$, the phase at the point of
the neutrino location,
$\psi=g\omega t(1-\beta/\beta_{0}\cos \phi)$,($g={\pm}1$), and the
amplitude, $B(t)$, which is a function of time. The wave speed in matter could be
less than the speed of light in vacuum $(\beta_0 \leq 1$), and $\phi$ is
the angle between the neutrino speed $\vec \beta$ and the direction of
the wave propagation. In the derivation of the Hamiltonian of eq.(2) terms
proportional to
${{1}\over{{\gamma}^{2}}}\ll{1}$,
$\gamma=(1-\beta^{2})^{-1/2}$,
and also an oscillating function of time in the diagonal part are neglected
(for details see [3-5]).

In order to study phenomenon of the parametric resonance of neutrino spin
oscillations in such a wave we suppose that the amplitude $B(t)$ is given by
\eq{B(t)=B(1+hf(t)), \ \ \ (|h|\ll 1),}
where $f(t)$ is an arbitrary function of time and $h$ is a small
dimensionless quantity of not fixed sign. It is convenient to introduce the
evolution operator which determines the neutrino state at time $t$
$$\nu(t)=U(t)\nu(0)$$
if the initial neutrino state was $\nu(0)$. Using the Hamiltonian of eq.(2)
we get the following equation for the evolution operator:
\eq{{\dot U}(t)=i[-{\tilde \rho}{\sigma_{3}}+(-E_{0}+\varepsilon f(t))
(\sigma_1 \cos \psi - \sigma_2 \sin \psi )]U(t),}
where c$\varepsilon=-E_{0}h$. In analogy with the case of the electromagnetic
wave with non - varying amplitude [3,4] the solution of eq.(4) can be written
in the form
\eq{U(t)=U_{\vec e_{3}}(t)U_{\vec l}(t)F(t),}
where
$U_{\vec e_{3}}(t)=\exp(i\sigma_{3}{{{\dot \psi}t}\over{2}})$
is the rotation operator around the axis
${\vec e}_{3}$ which is parallel with the direction of the neutrino
propagation, and
$U_{\vec l}(t)=\exp(i{\vec \sigma}{\vec l}t)$
is the rotation operator around the vector
${\vec l}=(-E_{0},0,{\tilde \rho}-{{\dot \psi}\over{2}})$.
It should be noted that the solution of eq.(4) for the case of constant
amplitude of the wave field $(\varepsilon=0)$ is given by the operator
$U_{0}(t)=U_{\vec e_{3}}(t)U_{\vec l}(t)$.

From (4) and (5) it follows that the equation for the operator $F(t)$ is
\eq{{\dot F}(t)=i\varepsilon
H_{\varepsilon}(t)F(t),}
where
$$H_{\varepsilon}(t)=({\vec \sigma}{\vec y}(t))f(t),$$

$${\vec y}=\big(1-2\lambda_{3}^{2}\sin^{2} \Omega t,
\ \lambda_{3}\sin 2\Omega t,\
2\lambda_{1}\lambda_{3}\sin^{2} \Omega t\big),$$
and the unit vector $\lambda$ is given by its components in the unit orthogonal
basis ${\vec \lambda}=
\lambda_{1}{\vec e}_{1}+\lambda_{2}{\vec e}_{2}+\lambda_{3}{\vec e}_{3}=
{{\vec l}\over{\Omega}}, \Omega=|{\vec l}|$. The detailed analysis of
evolution of the solution of eq.(6) can be found in [12]. Here we comment only
on the main steps. Using the smallness of $\varepsilon$ we expand the solution
of eq.(6) in powers of this parameter:
\eq{F=\sum_{k=0}^{\infty}\varepsilon^{k}F^{(k)},}
where
$F^{(0)}={\hat 1}$ is a unit matrix. For operators
$F^{(k)}$ the recurrent formula is straightforward:
\eq{F^{(k+1)}(t)=
i\int\limits_{0}^{t}H_{\varepsilon}(\tau)F^{(k)}(\tau)d\tau.}
Skipping further technical details to the first order in $\varepsilon$ we get
\eq{F(t)={\hat 1}+i\varepsilon({\vec \sigma}{\vec x}(t))+
O(\varepsilon^{2}),}
where
$${\vec x}(t)=
\int\limits\limits_{0}^{t} {\vec y}(\tau)f(\tau)d\tau.$$
Thus for the probability of neutrino conversion
$\nu_{i-}\leftrightarrow\nu_{j+}$
we get
\eq{
P_{ij}=
\big|<\nu_{+}|U_{\vec e_{3}}(t)U_{\vec l}(t)F(t)|\nu_{-}>\big|^{2}=}
$$\lambda_{1}^{2}\sin^{2}\Omega t+2\varepsilon \lambda_{1}
\big(x_{1}(t)\cos \Omega t+\lambda_{3}x_{2}(t)\sin \Omega t\big)
\sin \Omega t.$$
Note that $\Omega$ is nothing but the mean osillation frequency of the
neutrino system.

For further evaluation of solution of eq.(6) we have to specify the
form of the function $f(t)$. As it is mentioned above, the purpose
of this study is to examine the case when the parametric resonance for neutrino
oscillations in electromagnetic wave with varying amplitude could appear.
Having in mind a simple analogy ( see, for example, [10]) between oscillations
in neutrino system and oscillations of a classical pendulum [13], it is
reasonable to suppose that the principal parametric resonance appears when the
amplitude modulation function $f(t)$ is oscillating in time with frequency
approximately equal to the twice mean oscillation frequency of the system.
That is why we choose the function $f(t)$ to be
\eq{f(t)=\sin 2{\Omega}t}
and get for the neutrino oscillations probability
\eq{P_{ij}=\big[\lambda_{1}^{2}+\varepsilon
\lambda_{1}\lambda_{3}^{2}t+ {{\varepsilon \lambda_{1}}\over{\Omega}}
\big(1-{{\lambda_{3}^{2}}\over{2}}\big)\sin 2\Omega t\big]
\sin^{2}\Omega t.}
It follows that in the case $\lambda_{1}\varepsilon>0$ the second term
increases with increase of time $t$, so that the amplitude of the
neutrino conversion probability may becomes close to unity. This is the effect
of the parametric resonance in neutrino system in the electromagnetic wave
with modulated amplitude that may enhance neutrino oscillation amplitude even
for rather small values of the neutrino magnetic moment $\mu$ and strength
of the electromagnetic field and also when the system is
far away from the region of ordinary spin (or spin-flavour) neutrino resonance.

Let us consider the case when parameters of the neutrino system
are far beyond the region of the ordinary neutrino resonance. 
Supposing that $|l_{1}|=0.1|l_{3}|$. In this case the maximal neutrino
conversion probability, for non - varying $(h=0)$ amplitude
of the electromagnetic field, is indeed small,

\eq{P_{ijmax}(h=0)={{l_{1}^{2}}\over{l_{1}^{2}+l_{3}^{2}}}\ll 1.}
If the amplitude of the field is modulated in accordance with eq.(11)
the estimation for the critical time $t_{cr}$ for which the probability
$P_{ij}$
could become close to unity for an arbitrary values of the neutrino mixing
angle, masses, energy, and density of matter gives

\eq{t_{cr}\approx{{1}\over{\varepsilon n_{1}}}.}
It means that the parametric resonance enhancement of the neutrino
oscillations occurs after neutrinos travel a distance under the 
influence of electromagnetic wave with modulated 
amplitude. For the further numerical estimates me chose the strength 
of the field $B\approx 10^{3}$ G, the value of the neutrino magnetic 
moment $\mu \approx 10^{-10}\mu_{B}$ ($\mu_{B}$ is the Borh magneton),
and also suppose that neutrino is propagating in the direction 
opposite to one of the electromagnetic wave. Thus, from (14) we obtain, 
for the small time variations of the field amplitude 
(taking, for instance $|h|=0.1$), that the conversion probability
could attain the value $P_{ij}\approx 1$ after the neutrino has 
travelled a distance 

\eq{x_{min}=10^{-12}m.}
%


Let us consider now a system of two helicity neutrino states with 
mixing $\nu=(\nu_{+},\nu_{-})$. In general the neutrino components 
can belong to different flavours. The evolution of such a system in 
matter under the influence of transversal magnetic field  
is described by the Schr{\"o}dinger equation:

\eq{i{{\partial \nu}\over{\partial t}}=H\nu}
with Hamiltonian

\eq{H=-\mu B(t)\sigma_{1}+{\tilde \rho}\sigma_{3},
{\tilde \rho}={{\Delta m^{2}}\over{4E}}A-{{V_{eff}}\over{2}}}
Now we are going to discuss a possibility of parametric resonance 
amplification that could exist under the influence of constant in 
time but periodically varying in space magnetic field. We consider a 
special case when the strength of the field varies along the neutrino
path and the 
dependence on the neutrino coordinate $t$ is given by,

\eq{B(t)=\left\{
\begin{tabular}{l l}
$B_{1}$, & $0\leq t <T_{1}$ \\
$B_{2}$, & $T_{1}\leq t <T_{1}+T_{2}$ \\
\end{tabular}
\right.}
here $B_{1}$ and $B_{2}$ are constants. The field strength is periodic 
"castle wall"
function with period $T=T_{1}+T_{2}$: $B(t+T)=B(t)$.
The effective Hamiltonian is 
also a periodic function of time with the same period $T$:

\eq{H(t+T)=H(t),}
and

\eq{H(t)=\left\{
\begin{tabular}{l l}
$H_{1}$, & $0\leq t <T_{1}$ \\
$H_{2}$, & $T_{1}\leq t <T_{1}+T_{2}$ \\
\end{tabular}
\right.}
$H_{1}$, $H_{2}$ are the constant operators.
It should be noted that the similar approach to neutrino oscillations 
in the 
case of periodic step - function ("castle wall") density was 
described in details in the first paper of ref.[10]

Let us define the evolution operators for the two time intervals 
$(0,T_{1})$ and $(T_{1},T)$:

\eq{U_{a}=\exp(-iH_{a}T_{a}), \quad a=1,2.}
Then the operator giving the neutrino evolution for the whole period 
$T$ is

\eq{U_{T}=U_{2}U_{1}.}
For the further consideration it is convenient to introduce the unit 
vectors

\eq{{\vec n}_{a}=
-{{1}\over{\omega_{a}}}(E_{a},0,-{\tilde \rho})=
(\sin 2\theta_{a},0,-\cos 2\theta_{a}), \quad a=1,2,}
where $E_{a}=\mu B_{a}$, $\omega_{a}=
\sqrt{{\tilde \rho}^{2}+E_{a}^{2}}$ and
$\theta_{a}$ are the effective neutrino
mixing angles accounting for interaction 
with matter and magnetic field. Then for effective Hamiltonians we 
can write

\eq{H_{a}=\omega_{a}({\vec \sigma}\cdot{\vec n}_{a}),}

As it is easy to see, the eigenvalues of $H_{a}$ are 
${\pm \omega}_{a}$. 
Using eqs.(18-24) we get the 
followig expressions for the neutrino evolution operator over the 
whole period $T$:

\eq{U_{T}=Y-i({\vec \sigma}\cdot{\vec X})=
\exp(-i({\vec \sigma}\cdot{\vec n}_{X})\Phi),}
where
\eq{Y=c_{1}c_{2}-({\vec n}_{1}\cdot{\vec n}_{2})s_{1}s_{2},}
\eq{{\vec X}=s_{1}c_{2}{\vec n}_{1}+s_{2}c_{1}{\vec n}_{2}-
({\vec n}_{1}\times{\vec n}_{2})s_{1}s_{2},}
\eq{\Phi=\arcsin Y=\arccos X,}
\eq{{\vec n}_{X}={{\vec X}\over{X}}, \quad X=|\vec X|.}
and
\eq{s_{a}=\sin \phi_{a}, \quad c_{a}=\cos \phi_{a},
\quad \phi_{a}=\omega_{a}T_{a}, \quad a=1,2.}

From (26),(27) it follows that 

\eq{{\vec n}_{1}{\vec n}_{2}={{1}\over{{\omega}_{1}{\omega}_{2}}}
(B_{1}B_{2}+{\tilde \rho}^{2})=\cos 2({\theta}_{1}-{\theta}_{2}),}

\eq{{\vec n}_{1}\times{\vec n}_{2}={{1}\over{{\omega}_{1}{\omega}_{2}}}
(0,{\tilde \rho}(B_{1}-B_{2}),0)=
(0,\sin 2({\theta}_{1}-{\theta}_{2}),0).}
It should be noted also that $Y^{2}+{\vec X}^{2}=1$ due to unitarity of 
the operator $U_{T}$. The 
vector ${\vec X}$ can be expressed in its components as

\eq{{\vec X}=\Big(-
{{s_{1}c_{2}E_{1}}\over{\omega_{1}}}-{{s_{2}c_{1}E_{2}}\over{\omega_{2}}},
{\tilde \rho}{{s_{1}s_{2}}\over{\omega_{1}\omega_{2}}}(E_{2}-E_{1}),
{\tilde \rho}\big({{s_{1}c_{2}}\over{\omega_{1}}}+
{{s_{2}c_{1}}\over{\omega_{2}}}\big)
\Big).}

Let us consider the case when neutrino path in matter and periodic 
"castle wall" magnetic field is exactly equal to a whole number of 
periods, so that $t=nT, n\in N$. 
The neutrino evolution operator for $n$ periods is 
equal to $U_{T}$ in the $n$th power. 
Assume that initially the two - level 
neutrino system concerned occupies the state ${\nu}_{-}$.

\eq{U_{nT}=\exp(-i({\vec \sigma}\cdot{\vec n}_{X})n\Phi).}

Assume that initially $(t=0)$ the two - level neutrino system involved
occupies the state $\nu_{-}$. The transition probability to the state
$\nu_{+}$ after a time $t>0$ is determined by the evolution operator
$U_{T}$:

\eq{P(t)=|<\nu_{+}|U(t)|\nu_{-}>|^{2}.}
For the case $t=nT$ using eqs.(34),(35) we get 

\eq{P(nT)=
{{X_{1}^{2}+X_{2}^{2}}\over{X_{1}^{2}+X_{2}^{2}+X_{3}^{2}}}
\sin^{2}(n\Phi)=
{{X_{1}^{2}+X_{2}^{2}}\over{X_{1}^{2}+X_{2}^{2}+X_{3}^{2}}}
\sin^{2}(\Phi{{t}\over{T}}).}
This expression coincides with one obtained in the first paper 
of ref.[10] 
for neutrino 
oscillation probability in step - function profile of matter density. 
However, it is obvious that in our case
$X_{i}$ and $\Phi$ have different meaning.

Expression (36) also looks like the neutrino oscillations probability 
for the case of not varying in space (non-periodic) magnetic field. However,
there is an important distinction from the case of non-periodic magnetic 
field. The pre-sine factor in the latter case equals 
${{{{\mu}B}^{2}}\over{{{\mu}B}^{2}+{\tilde \rho}^{2}}}$
which is small if the value ${{\mu}B}$ is rather small. 
Contrary to the case of  
non-periodic magnetic field, in the case of $B_{1} \not= B_{2}$ 
the pre-sine factor need not 
be small even if $\sin^{2}2\theta_{a}$ are small. 
For particular choice of parameters the pre-sine term in eq.(36) 
becomes equal to unity. It is this case which we call the parametric 
enhancement of neutrino oscillations in periodic "castle wall" magnetic 
field. The resonance condition is 

\eq{X_{3}^{2}={\tilde \rho}^{2}
\big(
{{s_{1}c_{2}}\over{\omega_{1}}}+{{s_{2}c_{1}}\over{\omega_{2}}}
\big)^{2}=0,}

Now let us consider the case when the neutrino oscillations can be 
considerably increased due to the effect of parametric  resonance in
"castle wall" magnetic field. 
Assume 
that the magnetic field is rather small.  

\eq{(\mu B_{a})^{2}=E_{a}^{2}\ll \Delta^{2},}
Then from the resonance condition 
of eq.(37) it follows that

\eq{\phi_{1}+\phi_{2}=\pi k, \quad k\in N.}
If we introduce the mean oscillation frequency as

\eq{{\bar \omega}={{\omega_{1}T_{1}+\omega_{2}T_{2}}\over{T}},}
then the resonance condition can be rewritten in the form 

\eq{\Omega={{2{\bar \omega}}\over{k}},}
where $\Omega={{2\pi}\over{T}}$ is the magnetic field variation 
frequency ($T=T_{1}+T_{2}$). The last formula again reveals the 
well known  feature of the parametric resonance 
in an oscillating system: the 
resonance occurs when the doubled frequency, $2{\bar \omega}$, 
is equal to the frequency of 
a parameter variation multiplied by an integer number. 

Consider the transition probability, eq.(36), in the case of 
the parametric resonance (39). 
Thus from (33) for this case we get

\eq{{\vec X}^{2}_{१}={{1}\over{{\tilde \rho}^{2}}}
(s_{1}^{2}E_{1}^{2}+s_{2}^{2}E_{2}^{2}+2s_{1}E_{1}s_{2}E_{2}(-1)^{k}).}
Following the analysis performed in the first paper of ref.[10], 
one can show that the optimal 
conditions
for having transition probability are realised when

\eq{\phi_{a}={{\pi}\over{2}}+\pi k_{a}, \quad a=1,2,}
and $k_{1}+k_{2}\geq 0$ . Accounting (38) and (43) from (42) 
it follows that 

\eq{|\vec X|_{१}=
\left|{{E_{1}-E_{2}}\over{\tilde \rho}}\right|\ll 1.}
Thus, for the transition probability (36) in the case of 
parametric resonance we get 

\eq{P(nT)=
\sin^{2}\left(n{{(E_{1}-E_{2})}\over{\tilde \rho}}\right)=
\sin^{2}(2n(\theta_{1}-\theta_{2})).}
In derivation of the last expression we use the conditions 
$\theta_{a}\ll 1$.
The value of ${\tilde \rho}$ being equal to ${{\pi k}\over{T}}$, the 
formula (45) can be rewritten in the form

\eq{P(nT)=
\sin^{2}\left({{(E_{1}-E_{2})}\over{\pi k}}t\right)}
It follows that the parametric enhancement of neutrino 
oscillations becomes most 
sufficient when $k=1$. The similar effect exists for the 
parametric resonance in 
mechanical oscillating system that underlines the analogy mentioned above.

Let us perform numerical estimates of possible 
effect of the parametric 
resonance amplification of neutrino oscillations 
in "castle - wall" magnetic field. We 
assume for simplicity that $T_{1}=T_{2}=D$ and neglect matter effects. 
From the resonance 
condition (41) follows that

\eq{D={{2\pi kE}\over{\Delta m^{2}A}}.}
To obtain a maximal effect we chose $k=1$. 
Then, for the particular choice of the 
neutrino characteristics $\Delta m^{2}=0.1 eV^{2}$,
$E=0.1 ŒeV$, $\theta_{vac}=0$, we get $D=1m$. 
If the magnetic field profile corresponds to a chain of solenoids 
with opposite direction of electric currents, 
that results in $B_{1}=-B_{2}=B$ in terms of magnetic field direction, 
then for the oscillation probability we
get 

\eq{P(nT)=\sin^{2}(4n\theta),}
where $2\theta=\mu B/({{\Delta m^{2}}\over{4E}})$
For the neutrino transition magnetic moment $\mu=10^{-10}\mu_{B}$
and the strength of magnetic field $B=10^{7}$ G, we obtain   
$2\theta \approx 2.3 \cdot 10^{-5}$. Thus, the probability is given by

\eq{P(nT)=\sin^{2}(4.6\dot 10^{-5}n),}
The probability attains the value $P(nT)\approx 0.1$
(10\% of the total number of neutrinos are converted) 
when the number of periods equals to 7000.

The authors are thankful to Evgeny Akhmedov and Sergey Petcov 
for useful discussions. One of us (A.S.) should like to thank Mario Greco 
for invitation to Valee d'Aoste and to thank all the organizers of
this important meeting for hospitality.


\end{document}